\documentclass[traditabstract]{aa}

\usepackage{graphicx}
\usepackage{txfonts}
\usepackage{xcolor}
\usepackage{comment}
\usepackage{array}
\usepackage{natbib}
\bibpunct{(}{)}{;}{a}{}{,}
\usepackage{hyperref}
\hypersetup{colorlinks=true,
            citecolor=blue,
            linkcolor=blue}
\usepackage{color}
\def \vv#1{\vec{#1}}
\def \be {\begin{equation}}
\def \ee {\end{equation}}

\def\crm{\cr\noalign{\medskip}}

\def\atm {a}
\def\Va {V}
\def\DV {\Delta V}
\def\AT {\Lambda}
\def\VT {\Theta}
\def\SA {\Omega}
\def\ma {{\cal M}}
\def\inW {W}
\def\osdot {\star}
\def\Sc {{\cal S}_\osdot}
\def\cosSl {\ur \cdot  \ur'}
\def\cosSo {\ur \cdot  \ur_\osdot}
\def\cosSol {\ur' \cdot  \ur_\osdot}
\def\Er {K}
\def\Eo {E}
\def\Po {\dot \Eo}
\def\Pt {\dot U}

\def\ii {\mathrm{i}}

\def\Gc {{\cal G}}
\def\Ms {m_\osdot}
\def\ms {m}
\def \om {\omega}
\def \up {\upsilon}
\def \vpi {\varpi}

\def \cT {x}
\def \thetao {\varepsilon}

\def\xii {\hat x_1}
\def\xj {\hat x_2}
\def\xk {\hat x_3}

\def \vr {\vv{r}}
\def \ur {\hat \vr}

\def \vk {\vv{k}}

\def \vp {\vv{p}}
\def \vq {\vv{q}}
\def \vw {\boldsymbol{\om}}
\def \vs {\vv{s}}
\def \vu {\vv{r}}
\def \vg {\vv{a}}
\def \vL {\vv{L}}
\def \vF {\vv{F}}
\def \vG {\vv{G}}

\def \vT {\vv{T}}
\def\TI {{\bf \cal I}}
\def\SR {{\bf \cal S}}

\def \aZwpZn {a(0)}
\def \aZwpUn {a(n)}
\def \aZwpDn {a(2n)}
\def \aZwpTn {a(3n)}
\def \aUwpZn {a(\om)}
\def \aUwpUn {a(\om+n)}
\def \aUwmUn {a(\om-n)}
\def \aUwpDn {a(\om+2n)}
\def \aUwmDn {a(\om-2n)}
\def \aUwpTn {a(\om+3n)}
\def \aUwmTn {a(\om-3n)}
\def \aDwpZn {a(2\om)}
\def \aDwpUn {a(2\om+n)}
\def \aDwmUn {a(2\om-n)}
\def \aDwpDn {a(2\om+2n)}
\def \aDwmDn {a(2\om-2n)}
\def \aDwpTn {a(2\om+3n)}
\def \aDwmTn {a(2\om-3n)}

\def \bZwpUn {b(n)}
\def \bZwpDn {b(2n)}
\def \bZwpTn {b(3n)}
\def \bUwpZn {b(\om)}
\def \bUwpUn {b(\om+n)}
\def \bUwmUn {b(\om-n)}
\def \bUwpDn {b(\om+2n)}
\def \bUwmDn {b(\om-2n)}
\def \bUwpTn {b(\om+3n)}
\def \bUwmTn {b(\om-3n)}
\def \bDwpZn {b(2\om)}
\def \bDwpUn {b(2\om+n)}
\def \bDwmUn {b(2\om-n)}
\def \bDwpDn {b(2\om+2n)}
\def \bDwmDn {b(2\om-2n)}
\def \bDwpTn {b(2\om+3n)}
\def \bDwmTn {b(2\om-3n)}

\def \kl {{k'}}
\def \sumk {\sum_{k=-\infty}^{+\infty}}
\def \sumkl {\sum_{\kl=-\infty}^{+\infty}}

\def \tu {\tau}

\def \sq {\sqrt{1-e^2}}

\def \At {{\cal K}}

\def \az {a(kn)}
\def \au {a(\om-kn)}
\def \ad {a(2\om-kn)}
\def \bz {b(kn)}
\def \bu {b(\om-kn)}
\def \bd {b(2\om-kn)}
\def \Xtz {X_k^{-2,0} X_k^{-3,0}}

\def \Xtdp {X_k^{-2,2} X_k^{-3,2}}
\def \Xtdm {X_k^{-2,-2} X_k^{-3,-2}}
\def\ep {\mathrm{e}}

\def\tl {t'}

\def\iinf {\int}
\def\spi{}
\def\sqpi{}

\def\figpath{}
\def\bibpath{}
\def \llabel#1{\label{#1}}
\begin{document} 

\title{Spin and orbital dynamics of planets undergoing thermal atmospheric tides using a vectorial approach}

\titlerunning{Thermal atmospheric tides using a vectorial approach}

\authorrunning{E. F. S. Valente \& A. C. M. Correia}
\author{
Ema F. S. Valente\inst{1}%\orcid{0000-0002-9991-8270}
\and
Alexandre C. M. Correia\inst{1,2}%\orcid{0000-0002-8946-8579} 
}

\institute{
CFisUC, Departamento de F\'isica, Universidade de Coimbra, 3004-516 Coimbra, Portugal
\and 
IMCCE, Observatoire de Paris, PSL Universit\'e, 77 Av. Denfert-Rochereau, 75014 Paris, France
}

\date{Received; accepted To be inserted later}

  \abstract{
Earth-mass planets are expected to have atmospheres and experience thermal tides raised by the host star. These tides transfer energy to the planet that can counter the dissipation from bodily tides. Indeed, even a relatively thin atmosphere can drive the rotation of these planets away from the synchronous state. Here we revisit the dynamical evolution of planets undergoing thermal atmospheric tides. We use a novel approach based on a vectorial formalism, which is frame independent and valid for any configuration of the system, including any eccentricity and obliquity values. We provide the secular equations of motion after averaging over the mean anomaly and the argument of the pericenter, which are suitable to model the long-term spin and orbital evolution of the planet.
}


   \maketitle
%
%-------------------------------------------------------------------

\section{Introduction}
\label{Intro}

In general, Earth-mass planets are believed to be composed of a large rocky mantle covered by a thin atmospheric layer \citep[e.g.,][]{Komacek_Abbot_2019, Wordsworth_Kreidberg_2022}, as for the Earth, Venus, and Mars in the Solar System.
The molecules that make up the atmospheres can absorb part of the radiation they receive from the host star, giving rise to local temperature gradients, which in turn create pressure variations.
As a result, large-scale periodic mass redistribution inside the atmosphere occurs, while it attempts to return to an equilibrium state, known as thermal atmospheric tides.

Observations on Earth and Mars show that the pressure variations can be essentially decomposed in a diurnal and a semidiurnal tidal wave \citep[e.g.,][]{Withers_etal_2011, Auclair-Desrotour_etal_2017a}.
These oscillations are one of the most regular meteorological phenomena on Earth, and they are easily detectable by any station around the world \citep[e.g.,][]{Chapman_Lindzen_1970}.
The thermal inertia of the atmosphere introduces a delay between the stellar heating and the thermal response, which creates an asymmetry in the mass redistribution with respect to the substellar point.
As a consequence, the gravitational pull of the star exerts a torque on the atmosphere.
The diurnal wave has no net torque, but the semidiurnal wave gives rise to angular momentum exchanges within the atmosphere \citep[e.g.,][]{Correia_Laskar_2003JGR}.

Thermal atmospheric tides garnered special interest with the discovery of the retrograde rotation of Venus \citep[e.g.,][]{Carpenter_1964}. 
Because the atmosphere and the mantle are usually well coupled by friction at the surface, the variations in the angular momentum of the atmosphere are then transferred to the mantle of the planet, modifying its spin.
To explain the peculiar rotation of Venus, \citet{Gold_Soter_1969} thus proposed that it can be the result of a balance between bodily tides (i.e., gravitational tides raised in the mantle), which drive the planet toward synchronous rotation, and thermal tides, which drive it away.
This effect is negligible on Earth, because it is too distant from the Sun; however, for Earth-like planets in the habitable zone of K-dwarf stars, thermal tides can lead to asynchronous equilibria similar to that of Venus \citep{Leconte_etal_2015}.

Formation studies predict that Earth-mass planets are common around main sequence dwarf stars \citep[e.g.,][]{Schlecker_etal_2021}.
Indeed, there is already a large population of detected low-mass exoplanets, and their number is likely to grow rapidly \citep[e.g.,][]{Winn_2018}.
These planets are found in a wide range of orbital configurations, from very eccentric to compact multibody resonant orbits \citep[e.g.,][]{Borucki_etal_2013}.
We thus need a correct modeling of their spins and orbital dynamics.
The estimates for the tidal evolution of a planet are based on a general formulation of the tidal potential \citep[e.g.,][]{Kaula_1964}.
The classical expansion of this potential in elliptical elements depends on the chosen frame and introduces multiple index summations, which can lead to confusion and errors in the equations of motion.
Moreover, mistakes such as neglecting energy or momentum conservation considerations are more easily done \citep{Boue_Efroimsky_2019}.
For bodily tides, it has been shown that the equations of motion are more easily expressed in terms of angular momentum vectors \citep[e.g.,][]{Correia_Valente_2022}. 
Therefore, in this paper, we aim to also use these vectors to study the dynamics of a planet undergoing thermal tides.
This formalism is independent of the reference frame and allows us to simply add the contributions of multiple bodies in the system.

In Sect.~\ref{Maratmos}, we obtain the tidal potential of a planet whose atmosphere is excited through thermal forcing.
In Sect.~\ref{tbpwt}, we derive the equations of motion to study the spin and orbital evolution of a planet-star system under the action of thermal tides.
In Sect.\,\ref{secdyn}, we average the equations of motion over the mean anomaly and the argument of the pericenter, which provide simpler expressions that are easy to implement and suitable for long-term evolution studies.
In Sect.\,\ref{expe2} and Sect.\,\ref{planarcase}, we provide the equations of motion in the simplified cases of small eccentricity, and planar motion, respectively.
Finally, the last section is devoted to the conclusions.

\section{Thermal atmospheric tides}

\llabel{Maratmos}

We consider a planet with total mass $\ms$, which is composed of a perfectly spherical mantle with radius $R$, and a thin atmosphere with mass $\ma$.
We assume that the mantle is completely rigid, while the atmosphere can be deformed owing to the thermal forcing from the radiation of a nearby star.

\subsection{Atmospheric tidal potential}

The gravitational potential in a generic point of the space, $ \vr $, at a given time, $t$, generated by all the particles that compose the atmosphere is given by \citep[e.g.,][]{Goldstein_1950}
\be 
\Va (\vr,t) = - \Gc \int_\ma \frac{d \ma}{| \vr - \vr' |} 
\ , \llabel{220124a}
\ee 
where $ \vr' $ is the position of an atmosphere mass element $d \ma $. 
For a frame centered in the planet, we can expand Eq.\,(\ref{220124a}) in Legendre polynomials, $ P_\ell (x) $, as
\be 
\Va (\vr,t) = \sum_{\ell=0}^\infty V_{\ell} (\vr,t)
\ ,  \llabel{220209b} 
\ee
with
\be 
V_{\ell} (\vr,t) = - \frac{\Gc}{r} \int_\ma \left( \frac{r'}{r} \right)^{\ell} P_\ell (\cosSl) \, d \ma 
\ , \llabel{220630a}
\ee 
where $ \ur = \vr / r$ is the unit vector and $r = ||\vr||$ is the norm, and we assume $ r > r'$. 
For $\ell=0$ and $\ell=1$, we have $P_0 (\cosSl) = 1$ and $P_1 (\cosSl) = \cosSl$, respectively, and thus we can rewrite
\be
V_{0} (\vr,t) = - \frac{\Gc \ma}{r} 
 \llabel{220209d} 
\ee
and
\be
V_{1} (\vr,t) = - \frac{\Gc \ma}{r^2} \, \ur \cdot \vr_{\rm cm}
\ ,  \llabel{220209g} 
\ee
where 
\be 
\vr_{\rm cm} = \frac{1}{\ma} \int_\ma \vr' \, d \ma 
\ , \llabel{020103c}
\ee
is the center of mass of the atmosphere.
Furthermore, $V_0$ is the potential of a spherical symmetric atmosphere, which can be absorbed in the total potential of a point-mass planet.
For simplicity, we can set $\vr_{\rm cm}=0$, and thus $V_1 = 0$.
The resulting potential of a perturbed atmosphere, also known as tidal potential, is then given by the differential potential
\be
\DV (\vr,t) = \Va (\vr,t) - V_{0} (\vr,t)  - V_{1} (\vr,t) = \sum_{\ell=2}^\infty V_{\ell} (\vr,t) \ .
\ee
Moreover, if we assume $ r \gg r' $, we can keep only the term in $\ell=2$ (quadrupolar approximation), and finally get
\be 
\DV (\vr,t) \approx V_{2} (\vr,t) = - \frac{\Gc}{r^3} \int_\ma r'^2 P_2 (\cosSl) \, d \ma 
\ . \llabel{220630b}
\ee

\subsection{Deformation of the atmosphere}

For a frame attached to the planet, we can express $ \vr = (r, \theta, \phi) $ and $ \vr' = (r', \theta', \phi') $ in spherical coordinates.
Then
\be 
d \ma = \rho_\atm (\vr',t)  \, r'^2  \, d r' d \SA' 
\ , \llabel{220124b}
\ee 
where $ \rho_\atm (\vr',t) $ is the local density of the atmosphere, and 
\be
d \SA' = \sin \theta' d \theta' d \phi'
\ee is the differential solid angle.
Assuming a constant radius for the planet, $R$, we have
\be
\DV (\vr,t) =- \frac{\Gc}{r^3} \int\limits_{r' = R}^{+ \infty} \int\limits_{\phi'=0}^{2 \pi} \int\limits_{\theta'=0}^\pi
\rho_\atm (\vr',t) \, r'^4 P_2 (\cosSl) \, d r'  d \SA'
\ , \llabel{220630c}
\ee 
with
\be
\cosSl = \cos \theta \cos \theta' + \sin \theta \sin \theta' \cos (\phi - \phi')
\ . \llabel{220125a}
\ee
For terrestrial planets, the height of the atmosphere, $H$, is usually negligible compared to the radius of the planet\footnote{More generally, we have $ p(R+z) \approx p(R) \, \exp (-z / H_0) $, where $H_0$ is the ``scale height''. In the case of the Earth, $H_0 \sim 8 $~km, and because the pressure decreases exponentially, 95\% of the mass of the gas is contained in $H \sim 3 H_0 \sim 24$~km $\ll R_\oplus$ \citep[e.g.,][]{Chapman_Lindzen_1970}.}, that is
\be 
R \le r' \le R + H \ , 
\quad \mathrm{and} \quad 
\frac{H}{R} \ll 1
\ . \llabel{220124c}
\ee 
Thus, we can take a thin layer approximation and get
\be
\DV (\vr,t) = - \Gc R \left( \frac{R}{r} \right)^3 \int\limits_{r' = R}^{R+H} \int\limits_{\phi'=0}^{2 \pi} \int\limits_{\theta'=0}^\pi
 \rho_\atm (\vr',t) P_2 (\cosSl)  \, d r ' d \SA'
\ . \llabel{220125d}
\ee  
We further assume that the self-gravity fluctuations in the atmosphere can also be neglected \citep[e.g.,][]{Cowling_1941}.
Then, we use the hydrostatic equilibrium equation,
\be
d p = - \rho_\atm (\vr',t) g (\vr') \, d r'
\ , \llabel{220125b}
\ee
where $p (\vv r', t)$ is the local pressure and $g( \vv r')$ is the local gravity, to eliminate the integral in height 
\be 
\DV (\vr,t) = - \frac{\Gc R}{g} \left( \frac{R}{r} \right)^3 \int\limits_{\phi'=0}^{2 \pi} \int\limits_{\theta'=0}^\pi p_s (\theta', \phi',t)  P_2 (\cosSl) \, d \SA'
\ , \llabel{chp02.20} 
\ee  
and where $p_s (\theta', \phi',t) = p (R, \theta', \phi',t) $ is the surface pressure. 
We also assume that $ p (R+H, \theta', \phi',t) = 0 $, and that $ g (\vr') = g(R) = g $ is constant.
In a very general formulation, the surface pressure depends on the amount of energy per unit area received from the star, that is, the insolation $\inW (\theta', \phi',t)$, 
\be
p_s (\theta', \phi',t) = F \left[ \inW (\theta', \phi',t) \right]
\ , \llabel{220207a}
\ee
with $F$ being an operator that depends on the composition and the physical properties of the atmosphere. 
The insolation $W$ can be seen as a forcing function.
For a measurement at the top of the atmosphere, 
the insolation is given by \citep[e.g.,][]{Ward_1974}
\be
 W (\theta', \phi',t) =  \Sc \left( \frac{a}{r_\osdot } \right)^2 \left\{ 
  \begin{array}{c c}
     \cosSol  \ , & \cosSol > 0 \\\\
     0  \ , & \cosSol  \le 0 \\
   \end{array} 
 \right. \ ,
\llabel{010611b} 
\ee
where $\Sc = L_\osdot / (4\pi a^2)$ is the ``solar'' constant, $L_\osdot $ is the luminosity of the star, $ \vr_\osdot$ is the position of the star with respect to the center of mass of the planet (which is a function of the time),  $a$ is the semimajor axis of the star-planet orbit, and 
\be
\cosSol  = 
 \cos \theta' \cos \theta_\osdot  + \sin \theta' \sin \theta_\osdot  \cos (\phi' - \phi_\osdot ) 
\ . \llabel{220125e} 
\ee
We can expand expression (\ref{010611b}) in Legendre polynomials, as
\be
 W (\theta', \phi',t) = \sum_{n=0}^\infty  W_n \, \VT_n (\vv r', t) 
\ , \llabel{010613a} 
\ee 
with
\be
\VT_n (\vv r, t ) 
= \left( \frac{a}{r_\osdot} \right)^2 P_n \left( \cosSo \right) 
\ , \llabel{220207b}
\ee
and
\be 
 W_n = \Sc \frac{2 n + 1}{2}  \int_0^1 x P_n (x) \, d x 
\ . \llabel{010618c} 
\ee
The insolation variations can also be expressed in the frequency domain by performing a Fourier transform
\be
 W (\theta', \phi',t) =  \sum_{n=0}^\infty \sqpi \iinf  W_n \,  \hat \VT_n (\vv r', \sigma)  \, \ep^{\ii \sigma t} \, d \sigma 
\ , \llabel{230227a} 
\ee 
with
\be
\hat \VT_n (\vv r, \sigma) = \sqpi \iinf   \VT_n (\vv r, t ) \, \ep^{-\ii \sigma t} \, d t 
\llabel{230227c} 
\ee
Since the Legendre polynomials form an orthogonal basis, any other quantity with the same symmetry as the insolation, such as the surface pressure variations (Eq.\,(\ref{220207a})), can also be expanded in a similar way.
Therefore, we write
\be
\begin{split}
p_s (\theta', \phi',t) & 
= \sum_{n=0}^\infty \sqpi \iinf  \hat p_n(\sigma) \,  \hat \VT_n (\vv r', \sigma)  \, \ep^{\ii \sigma t} \, d \sigma \ , \llabel{220125g}  
\end{split}
\ee
where $ \hat p_n $ are the coefficients of order $ n $ of the decomposition of the surface pressure in Legendre polynomials.
They can also be expressed in terms of spherical harmonics $Y_\ell^m$ and their complex conjugate $\bar Y_\ell^m$ \citep[e.g.,][]{Abramowitz_Stegun_1972}
\be
P_\ell (\cosSl) = \frac{4 \pi}{2 \ell +1} \sum_{m = -\ell}^{\ell} Y_\ell^m (\theta, \phi) \bar Y_\ell^m (\theta', \phi') 
\ . \llabel{220125h} 
\ee
Then, by replacing (\ref{220125g}) and (\ref{220125h}) in expression (\ref{chp02.20}) we get
\be 
\begin{split}
\DV (\vr,t) = - \frac{\Gc R}{g} \left( \frac{R}{r} \right)^3 \frac{4 \pi}{5} \sum_{m = -2}^{2} Y_2^m (\theta, \phi) \sum_{n=0}^\infty \sqpi \iinf  \hat p_n(\sigma) \, \times \\
\int\limits_{\phi'=0}^{2 \pi} \int\limits_{\theta'=0}^\pi  \hat \VT_n (\vv r', \sigma) \bar Y_2^m (\theta', \phi') \, d \SA' \, \ep^{\ii \sigma t} \, d \sigma
\ , \llabel{220125i} 
\end{split}
\ee  
and, using the orthogonality formula 
\be
\int\limits_{\phi'=0}^{2 \pi} \int\limits_{\theta'=0}^\pi P_n \left( \cosSol \right) \bar Y_2^m (\theta',\phi')  \, d \SA'  = \frac{4 \pi}{5} \bar Y_2^m (\theta_\osdot, \phi_\osdot) \, \delta_{n ,2} 
\ , \llabel{220207c} 
\ee
where $ \delta_{n,2} $ is the Kronecker delta, we finally get
\be 
\DV (\vr,t) = - \frac{\Gc R}{g} \left( \frac{R}{r} \right)^3 \frac{4 \pi}{5} \sqpi \iinf  \hat p_2(\sigma) \, \hat \VT_2 (\vv r, \sigma) \, \ep^{\ii \sigma t} \, d \sigma
\ . \llabel{220210b} 
\ee 

The mass distribution inside the planet is usually better characterized by its inertia tensor.
Then, we can rearrange Eq.\,(\ref{220210b}) as
\be
\DV (\vr_B ,t ) = \frac{3 \Gc}{2 r^3} \, \ur_B \cdot \TI_B (t) \cdot \ur_B 
 \ , \llabel{220210dB}
\ee
with
\be 
\TI_B (t) 
= \sqpi \iinf  \hat p_2(\sigma) \, \hat \AT_B (\sigma) \, \ep^{\ii \sigma t} \, d \sigma
\ , \llabel{220210e} 
\ee
\be
\hat \AT_B (\sigma) = \sqpi \iinf  \AT_B (t) \, \ep^{-\ii \sigma t} \, d t \ ,
\llabel{210929eB}
\ee
and
\be 
\AT_B (t) =  -  \frac{4 \pi R^4}{5 g}  \left( \frac{a}{r_\osdot} \right)^2 \, \left( \ur_{\osdot B}  \, \ur_{\osdot B}^T - \frac13  \mathbb{I} \right) 
\ , \llabel{220210fB} 
\ee
where $^T$ denotes the transpose, $\mathbb{I}$ is the identity matrix, $\TI (t) $ is the atmosphere inertia tensor that accounts for the departure of the mass distribution from a sphere (Eq.\,(\ref{121026c})), while $\AT (t)$ is a forcing tensor that depends on time through $\vr_{\osdot}$.
The subscript $_B$ is there to remind us that Eqs.\,(\ref{220210dB})$-$(\ref{220210fB}) were obtained in the body frame, that is, in a frame attached to the planet.

\subsection{Deformation in an arbitrary frame}

It is also possible to compute the deformation of the atmosphere in an arbitrary frame, $\TI(t)$.
Following \citet{Correia_Valente_2022}, we let $\SR$ be the rotation matrix that allows us to convert any vector $\vu_B$ in a frame attached to the planet into another frame $\vu$, such that $\vu = \SR \, \vu_B$, and
\be
\TI (t) = \SR(t) \, \TI_B(t) \, \SR(t)^T 
\ . \llabel{200116c}
\ee
Then, the atmospheric tidal potential (Eq.\,(\ref{220210dB})) becomes
\be
\DV (\vr ,t ) = \frac{3 \Gc}{2 r^3} \, \ur \cdot \TI (t) \cdot \ur 
 \ . \llabel{220210d}
\ee
Similarly, we can also compute the forcing tensor (Eq.\,(\ref{220210fB})) in the new frame as
\be
\AT (t) = \SR(t) \, \AT_B(t) \, \SR(t)^T  
 = -  \frac{4 \pi R^4}{5 g}  \left( \frac{a}{r_\osdot} \right)^2 \, \left( \ur_{\osdot}  \, \ur_{\osdot}^T - \frac13  \mathbb{I} \right) 
\ . \llabel{220210f} 
\ee

We note that the expression of $\TI_B(t)$ (Eq.\,(\ref{220210e})) also corresponds to the convolution product between $\tilde p_2 (t)$ and $\AT_B (t)$,
\be
\TI_B (t) = \Big[ \tilde p_2 * \AT_B \Big] (t) 
= \sqpi \iinf  \tilde p_2 (t-t') \,  \AT_B (t') \, d t' 
\ , \llabel{220630y} 
\ee
where 
\be
\tilde p_2 (t) = \sqpi \iinf \hat p_2(\sigma) \, \ep^{\ii \sigma t} \, d \sigma  \ .
\llabel{220701a} 
\ee
Then, in the frequency domain we have
\be
\begin{split}
\hat \TI (\sigma) =&
\sqpi \iinf \TI (t) \, \ep^{-\ii \sigma t} \, d t = 
\sqpi \iinf \SR(t) \, \TI_B(t) \, \SR(t)^T \ep^{-\ii \sigma t} \, d t \\
=& \spi \iint \tilde p_2(t-\tl) \, \SR(t) \, \AT_B(\tl) \, \SR(t)^T \ep^{-\ii \sigma t} \, d \tl d t \\
=& \spi \iint \tilde p_2(\tu) \, \SR(\tl+\tu) \, \AT_B(\tl) \,  \SR(\tl+\tu)^T \ep^{-\ii \sigma (\tl + \tu)} \, d \tl d \tu
\ . \llabel{200116d} 
\end{split}
\ee
Since $\SR(t)$ is a rotation matrix, we note that $\SR(\tl+\tu) = \SR(\tu) \SR(\tl)$, and thus
\be
\begin{split}
\hat \TI (\sigma) 
=& \spi \iint \tilde p_2(\tu) \, \SR(\tu) \, \AT(\tl) \, \SR(\tu)^T \ep^{-\ii \sigma (\tl + \tu)} \, d \tl d \tu \\
=& \sqpi \iinf  \tilde p_2(\tu) \, \SR(\tu) \, \hat \AT(\sigma) \, \SR(\tu)^T \ep^{-\ii \sigma \tu} \, d \tu
\ , \llabel{200116e} 
\end{split}
\ee
with
\be
\hat \AT (\sigma) = \sqpi \iinf  \AT (t) \, \ep^{-\ii \sigma t} \, d t 
\ . \llabel{210929e}
\ee

\begin{figure}
\begin{center}
\includegraphics[width=9.2cm]{\figpath 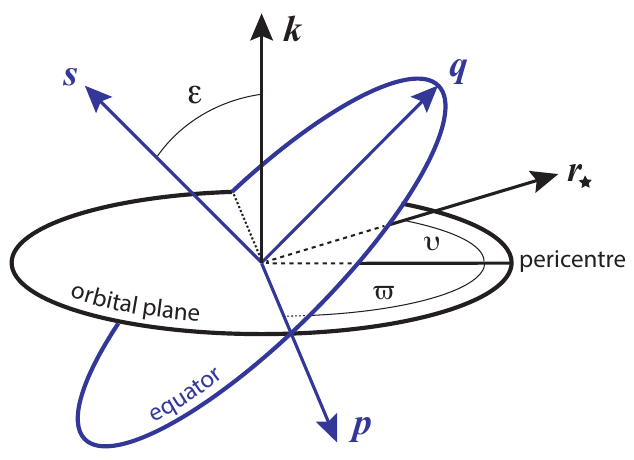} 
 \caption{Reference frame ($\vp,\vq,\vs$) and angles. We note that $\vk$ is a unit vector that is normal to the orbital plane, $\vs$ is a unit vector that is normal to the equatorial plane of the planet, and $\vp$ is a unit vector along the line of nodes of the two planes. The obliquity $\thetao$ is the angle between the two planes, $\vpi$ is the argument of the pericenter, and $\up$ is the true anomaly. \llabel{frames}  }
\end{center}
\end{figure}

\begin{table*}
\begin{center}
\begin{tabular}{|c|c|c| } \hline 
$ -\Delta \hat \AT (\sigma) \cos 2 \om \tu - \hat \AT_{12} (\sigma) \sin 2 \om \tu -\frac12 \hat \AT_{33} (\sigma) $ & $ -\Delta \hat \AT (\sigma) \sin 2 \om \tu + \hat \AT_{12} (\sigma) \cos 2 \om \tu $ & $ \hat \AT_{13} (\sigma) \cos \om \tu - \hat \AT_{23} (\sigma) \sin \om \tu $ \\ \hline
$ -\Delta \hat \AT (\sigma) \sin 2 \om \tu + \hat \AT_{12} (\sigma) \cos 2 \om \tu $ & $ \Delta \hat \AT (\sigma) \cos 2 \om \tu + \hat \AT_{12} (\sigma) \sin 2 \om \tu -\frac12 \hat \AT_{33} (\sigma) $ & $ \hat \AT_{13} (\sigma) \sin \om \tu + \hat \AT_{23} (\sigma) \cos \om \tu $ \\ \hline
$ \hat \AT_{13} (\sigma) \cos \om \tu - \hat \AT_{23} (\sigma) \sin \om \tu $ & $ \hat \AT_{13} (\sigma) \sin \om \tu + \hat \AT_{23} (\sigma) \cos \om \tu $ & $ \hat \AT_{33} (\sigma) $ \\ \hline
\end{tabular} 
\end{center}
\caption{Coefficients of the product matrix $ \SR(\tu) \, \hat \AT(\sigma) \, \SR(\tu)^T $ (Eq.\,(\ref{200116e})).  \llabel{tabSAS} }  
\end{table*}

At this stage, we need to adopt a specific frame to proceed, in order to express $\SR(\tu)$.
We adopt here a Cartesian reference frame $(\vp,\vq,\vs)$, such that $\vs$ is the spin axis, 
\be
\vp = \frac{ \vk \times \vs }{\sin \thetao} \ , 
\quad \vq = \frac{\vk - \cos \thetao \, \vs}{\sin \thetao}  \ , 
\quad \cos \thetao = \vk \cdot \vs \ ,
\llabel{221007a}
\ee
where $\vk$ is a unit vector that is normal to the orbital plane of the star, $\vp$ is aligned with the line of nodes between the equator of the planet and the orbital plane, and $\thetao$ is the angle between these two planes, also known as the obliquity (Fig.~\ref{frames}).
This particular frame is very useful to obtain the secular equations of motion (Sect.~\ref{sec_ttp}), as it allows us to decouple the rotational perturbations from the orbital ones \citep{Correia_2006}.
For a planet rotating about the $\vs$ axis with angular velocity $\om$, we thus have 
\be
\SR(\tu) = \left[\begin{array}{ccc} 
\cos \om \tu   &  -\sin \om \tu   & 0 \\
\sin \om \tu   & \cos \om \tu & 0 \\
0 & 0 & 1
\end{array}\right] 
\ . \llabel{200116f}
\ee
The result for the product $ \SR(\tu) \, \hat \AT(\sigma) \, \SR(\tu)^T$ is given in Table~\ref{tabSAS}.
For instance, for the $\hat I_{23} (\sigma)$ coefficient of the inertia tensor $\hat \TI (\sigma)$ (Eq.\,(\ref{121026c})), we then finally have
\be
\llabel{200117c} 
\begin{split}
\hat I_{23} (\sigma) & = \sqpi
\iinf  \tilde p_2(\tu) \left[ \hat \AT_{13} (\sigma) \sin \om \tu + \hat \AT_{23} (\sigma) \cos \om \tu \right] \ep^{-\ii \sigma \tu} \, d \tu  \\ 
& = 
\hat \AT_{13} (\sigma) \sqpi \iinf  \tilde p_2(\tu) \sin \om \tu \, \ep^{-\ii \sigma \tu} \, d \tu \\
& \quad \quad + \hat \AT_{23} (\sigma) \sqpi \iinf  \tilde p_2(\tu) \cos \om \tu \, \ep^{-\ii \sigma \tu} \, d \tu  \\
& = \frac12 \hat p_2 (\sigma-\om) \left[ \hat \AT_{23} (\sigma) - \ii \, \hat \AT_{13} (\sigma) \right] \\
&  \quad \quad + \frac12 \hat p_2 (\sigma+\om) \left[ \hat \AT_{23} (\sigma) + \ii \, \hat \AT_{13} (\sigma) \right] \ . 
\end{split}
\ee
Similarly, for the remaining coefficients of $\hat \TI (\sigma)$, we obtain
\be
\llabel{200117b} 
\begin{split}
\hat I_{13} (\sigma) & = \frac12 \hat p_2 (\sigma-\om) \left[ \hat \AT_{13} (\sigma) + \ii \, \hat \AT_{23} (\sigma) \right] \\
& \quad \quad + \frac12 \hat p_2 (\sigma+\om) \left[ \hat \AT_{13} (\sigma) - \ii \, \hat \AT_{23} (\sigma) \right] \ ,
\end{split}
\ee
\be 
\llabel{200117d} 
\begin{split}
\hat I_{12} (\sigma) & = \frac12 \hat p_2 (\sigma-2\om) \left[ \hat \AT_{12} (\sigma) + \ii \, \Delta \hat \AT (\sigma) \right] \\
&  \quad \quad
+ \frac12 \hat p_2 (\sigma+2\om) \left[ \hat \AT_{12} (\sigma) - \ii \, \Delta \hat \AT (\sigma) \right] \ ,
\end{split}
\ee
\be
\hat I_{33} (\sigma) =\hat p_2 (\sigma) \hat \AT_{33} (\sigma)
\ , \llabel{200117f} 
\ee
\be
\hat I_{22} (\sigma) = \Delta \hat I (\sigma) - \frac12 \hat I_{33} (\sigma)
\ , \llabel{200117g} 
\ee
\be
\hat I_{11} (\sigma) = - \Delta \hat I (\sigma) - \frac12 \hat I_{33} (\sigma)
\ , \llabel{200117h} 
\ee
with
\be
\llabel{200117e} 
\begin{split}
\Delta \hat I (\sigma) & = \frac12 \hat p_2 (\sigma-2\om) \left[ \Delta \hat \AT (\sigma) - \ii \, \hat \AT_{12} (\sigma) \right] \\
&  \quad \quad
+ \frac12 \hat p_2 (\sigma+2\om) \left[ \Delta \hat \AT (\sigma) + \ii \, \hat \AT_{12} (\sigma) \right] \ ,
\end{split}
\ee
and
\be
\Delta \hat \AT (\sigma) = \frac12 \left[ \hat \AT _{22} (\sigma) -  \hat \AT _{11} (\sigma) \right] \ .
\ee

\subsection{Surface pressure variations}

\llabel{tidalmodels}

The deformation of the atmosphere is given by the inertia tensor, $\hat \TI (\sigma)$, which is obtained from the second harmonic of the surface pressure, $\hat p_2 (\sigma)$, combined with the forcing tensor, $\hat \AT (\sigma)$ (Eq.\,(\ref{200116e})).
While the forcing tensor is well determined, as it only depends on the position of the star (Eq.\,(\ref{220210f})), the surface pressure is subject to many uncertainties, as it depends on the composition and the physical properties of the atmosphere.

In order to compute $\hat p_2 (\sigma)$, we need to adopt some dynamical model for the atmosphere.
The study of thermal tides has been initiated by the pioneer work of \citet{Siebert_1961} and \citet{Chapman_Lindzen_1970}.
They assumed that the atmosphere of the planet corresponds to an ideal gas obeying the perfect gas law, together with the conservation of mass and the Navier-Stokes equation.
These equations were then linearized around the equilibrium state.
\citet{Chapman_Lindzen_1970} studied the case of a fast rotating planet, such as the Earth, but for long-term evolution studies, the planet may encounter slow rotation regimes near synchronization, such as Venus \citep[e.g.,][]{Correia_Laskar_2001, Correia_Laskar_2003I}.
In the later case, it is important to take into account the effect of radiative losses \citep{Auclair-Desrotour_etal_2017a}.
Assuming a slowly rotating convective atmosphere, the hydrodynamic equations can be simplified drastically, since the Coriolis acceleration can be neglected.
The thin atmospheric layer close to the surface has a strongly negative temperature gradient \citep{Seiff_etal_1980}, which subjects this layer to convective instability \citep{Baker_etal_2000}.
As a result, gravity waves cannot propagate in the unstable region above the surface and we can set the Brunt-V\"ais\"al\"a frequency to approximately zero, which corresponds to the adiabatic temperature gradient.
Following \citet{Dobrovolskis_Ingersoll_1980}, thermal tides are thus solely considered to be generated by the average heat at the ground, $\hat J_0$, which finally gives the following for the second harmonic of the surface pressure variations \citep[][Eq.\,(166)]{Auclair-Desrotour_etal_2017a}:
\be
\hat p_2 (\sigma) = - \frac{\kappa \, \rho_0 \, \hat J_0}{\sigma_0 + \ii \sigma} 
\ , \llabel{220929a}
\ee
with $\kappa = 1- \gamma^{-1}$, where $\gamma$ is the adiabatic exponent,
$\rho_0 = \rho_a(0)$ is the mean surface density of the atmosphere, and
$\sigma_0$ is the radiative frequency of the atmosphere, which depends on its thermal capacity.
This theoretical estimation agrees incredibly well with empirical estimations derived from generic global climate circulation models \citep[][Fig.~1]{Leconte_etal_2015}.
Interestingly, it is also similar to the Love number of a Maxwell tidal model \citep[e.g.,][]{Correia_etal_2014, Auclair-Desrotour_etal_2019a}.
The minus sign in the expression of $\hat p_2 (\sigma)$ causes the pressure variations to lead the star, a phenomenon well documented for the Earth \citep[e.g.,][]{Chapman_Lindzen_1970}.

The $\hat p_2(\sigma)$ is a complex number, whose modulus gives the amplitude of the pressure variations and the argument gives the phase lag between the substellar point and the maximal deformation of the atmosphere.
It can be decomposed in its real and imaginary parts as
\be
\hat p_2 (\sigma) = a (\sigma) - \ii \, b (\sigma) 
\ , \llabel{210924d}
\ee
which is very useful when we write the secular equations of motion (see Sect.~\ref{sec_ttp}), because the imaginary part characterizes the atmosphere's viscous response.
We thus have
\be
a(\sigma) = -  \frac{\kappa \, \rho_0 \, \hat J_0 \, \sigma_0}{{\sigma_0}^2 + \sigma^2}
\ , \quad \mathrm{and} \quad
b(\sigma) = -  \frac{ \kappa \, \rho_0 \, \hat J_0 \, \sigma}{{\sigma_0}^2 + \sigma ^2}
\ . \llabel{210929z}
\ee
It is also important to note that $a (\sigma)$ is always an even function and $b (\sigma)$ is always an odd function.

\section{Two-body problem with thermal tides}

\llabel{tbpwt}

We consider a system composed by a planet and a star with masses $\ms$ and $\Ms$, respectively, on a Keplerian orbit.
The orbital angular momentum is given by
\be 
\vG = \beta \sqrt{\mu a (1-e^2)} \, \vk 
\ , \label{210804a}
\ee
where $a$ is the semimajor axis, $e$ is the eccentricity,
$\beta = \Ms \ms / (\Ms + \ms)$,
$\mu = \Gc (\Ms + \ms)$,
$\Gc$ is the gravitational constant,
and $\vk$ is the unit vector along the direction of $\vG$, which is normal to the orbit.
The star is a point mass object with luminosity $L_\osdot$.
The planet is composed by a completely rigid spherical mantle with radius $R$, and surrounded by a thin atmosphere that can be deformed under the action of thermal tides.
The planet rotates with angular velocity $\vw = \om \, \vs$, where $\vs$ is the unit vector along the direction of the spin axis.
The rotational angular momentum of the planet is given by
\be
\vL = C \vw + \TI \cdot \vw  
\ , \llabel{210804b}
\ee
where $C$ is the moment of inertia of the mantle together with the moment of inertia of an unperturbed atmosphere and
\be
\TI = 
\left[\begin{array}{rrr} 
I_{11}&  I_{12}& I_{13} \\
I_{12}&  I_{22}& I_{23} \\
I_{13}&  I_{23}& I_{33} 
\end{array}\right]
\label{121026c}
\ee
is the inertia tensor that accounts for the departure of the mass distribution in the atmosphere from a sphere (Eq.\,(\ref{200116c})).
In the absence of the stellar heating, we have $\TI=0$.
In general, the deformations in the atmosphere are very small with respect to the radius of the spherical mantle, $R$, and yield periodic changes in the moments of inertia, such that $\dot \TI \vw \ll C \dot \vw$ \citep[e.g.,][]{Frouard_Efroimsky_2018}.
Therefore, for simplicity, we can assume that $I_{ij} \ll C$ ($i,j=1,2,3$), and thus
\be
\vL \approx C  \vw  = C \om \, \vs
\ .  \llabel{151019b}
\ee

\subsection{Tidal force and torque}
\llabel{pmp}

The atmospheric tidal potential (Eq.\,\eqref{220210d}) creates a differential gravitational field around the planet given by
\be
\vg (\vr) = - \nabla_{\vr} \left[ \DV (\vr) \right] \ .
\ee
The star, with mass $\Ms$ and located at $\vr=\vr_\osdot$, interacts with this field, with a resulting tidal force
\be
\vF = \Ms \vg (\vr_\osdot) =  \vF_1 + \vF_2
\ , \llabel{170911d}
\ee
with

\be
 \llabel{170911b}
\begin{split}
\vF_1  & = 
\frac{15 \Gc \Ms}{r_\osdot^4}  \bigg[ \frac{I_{22}-I_{11}}{2} \big(\xj^2 - \frac15 \big)  
+ \frac{I_{33}-I_{11}}{2} \big(\xk^2 - \frac15 \big) \\ 
& \quad \quad + I_{12} \xii \xj + I_{13} \xii \xk + I_{23} \xj \xk \bigg] \, \ur_\osdot 
\end{split}
\ee
and
\be
\llabel{170911c}
\begin{split}
\vF_2 = &
 -  \frac{3 \Gc \Ms}{r_\osdot^4}  \bigg[ \big(I_{22}-I_{11}\big) \xj \, \vq + \big(I_{33}-I_{11}\big) \xk \, \vs  \\  & 
 + I_{12} (\xii \, \vq + \xj \, \vp)  + I_{13} (\xii \, \vs + \, \xk \, \vp) + I_{23} (\xj \, \vs + \xk \, \vq) \bigg]
\ ,
\end{split}
\ee
where $\ur_\osdot = \vr_\osdot / r_\osdot =  (\xii, \xj, \xk)$ is the unit vector. 
We decompose $\vF$ and all the following vectorial quantities in the frame $(\vp, \vq, \vs)$ (Fig.~\ref{frames}).
There is no loss of generality with this frame choice, because the vectors can always be expressed in another basis.
However, the frame $(\vp, \vq, \vs)$ greatly facilitates the computation of the inertia tensor $\TI$ (Sect.~\ref{hce}) since the position of the star does not depend on the planet's rotation rate:
\be
 \label{210804e}
\begin{split}
\xii & = \ur_\osdot \cdot \vp 
= \cos (\vpi+\up) \ , \\
\xj & = \ur_\osdot \cdot \vq =  \cos \thetao \sin (\vpi+\up) \ , \\
\xk & = \ur_\osdot \cdot \vs 
= - \sin \thetao \sin (\vpi+\up) \ ,
\end{split}
\ee
and
\be
\begin{split}
\dot \vr_\osdot \cdot \vp & =
\frac{- n a}{\sqrt{1-e^2}} \,  \big( \sin ( \vpi +  \up ) + e \sin \vpi \big) \ , \\
\dot \vr_\osdot \cdot \vq & =
 \frac{n a}{\sqrt{1-e^2}} \cos \thetao \, \big( \cos (\vpi + \up) + e \cos \vpi \big) \ , \\
\dot \vr_\osdot \cdot \vs & =
 \frac{-n a}{\sqrt{1-e^2}}  \sin \thetao \, \big( \cos (\vpi + \up) + e \cos \vpi \big) \ ,
\end{split}
\ee
where $n = \sqrt{\mu / a^3}$ is the mean motion, $\up$ is the true anomaly, and $\vpi$ is the argument of the pericenter (Fig.~\ref{frames}).

We follow the evolution of the system in an inertial frame because we can always project an inertial vector in a noninertial coordinate system.
For the orbital evolution, we thus obtain
\be
\ddot \vr_\osdot =  - \frac{\mu}{r_\osdot^2} \, \ur_\osdot + \frac{\vv{F}}{\beta} \ .  \llabel{151028c}
\ee
The first term is responsible for the Keplerian motion, while the second term corresponds to the orbital correction introduced by the tidal force, which is responsible for the modifications in the orbit and spin of the planet.
The evolution of the angular momentum vectors are computed from the tidal torque,
\be
\dot \vG = \vT = \vr_\osdot \times \vv{F} = \vr_\osdot \times \vF_2 
\ , \llabel{150626a}
\ee
and, due to the conservation of the total angular momentum,
\be
\dot \vL = - \dot \vG = - \vT
\ , \llabel{210805b}
\ee
with
\be
\vT = - \frac{3 \Gc \Ms}{r_\osdot^3}
%\small
\left[\begin{array}{c}  
\big(I_{33}-I_{22}\big) \xj \xk - I_{12} \xii \xk + I_{13} \xii \xj  + I_{23} (\xj^2 - \xk^2)  \crm
\big(I_{11}-I_{33}\big) \xii \xk + I_{12} \xj \xk + I_{13} (\xk^2 - \xii^2) - I_{23} \xii \xj  \crm
\big(I_{22}-I_{11}\big) \xii \xj + I_{12} (\xii^2 - \xj^2) - I_{13} \xj \xk + I_{23} \xii \xk
\end{array}\right] 
\ . \llabel{151028e}
\ee

\subsection{Expansion in Hansen coefficients}
\llabel{hce}

\begin{table*}
\begin{center}
\begin{tabular}{|r|c|c|c|c| } \hline 
$k$ & $X_k^{-2,0} (e) $ & $X_k^{-2,2} (e) $ & $X_k^{-3,0} (e) $ & $X_k^{-3,2} (e) $ \\ \hline
$-6$ & $\frac{1223}{320} e^6$ & $$ & $\frac{3167}{320} e^6$ & $$ \\
$-5$ & $\frac{1097}{384} e^5$ & $$ & $\frac{1773}{256} e^5$ & $$ \\
$-4$ & $\frac{103}{48} e^4 - \frac{129}{160} e^6$ & $-\frac{2}{45} e^6$ & $\frac{77}{16} e^4 + \frac{129}{160} e^6$ & $\frac{4}{45} e^6$ \\
$-3$ & $\frac{13}{8} e^3 - \frac{25}{128} e^5$ & $-\frac{27}{640} e^5$ & $\frac{53}{16} e^3 + \frac{393}{256} e^5$ & $\frac{81}{1280} e^5$ \\
$-2$ & $\frac{5}{4} e^2 + \frac{1}{6} e^4 + \frac{21}{64} e^6$ & $-\frac{1}{24} e^4 - \frac{3}{80} e^6$ & $\frac{9}{4} e^2 + \frac{7}{4} e^4 + \frac{141}{64} e^6$ & $\frac{1}{24} e^4 + \frac{7}{240} e^6$ \\
$-1$ & $e + \frac{3}{8} e^3 + \frac{65}{192} e^5$ & $-\frac{1}{24} e^3 - \frac{13}{384} e^5$ & $\frac{3}{2} e + \frac{27}{16} e^3 + \frac{261}{128} e^5$ & $\frac{1}{48} e^3 + \frac{11}{768} e^5$ \\
$0$ & $1 + \frac{1}{2} e^2 + \frac{3}{8} e^4 + \frac{5}{16} e^6$ & $$ & $1 + \frac{3}{2} e^2 + \frac{15}{8} e^4 + \frac{35}{16} e^6$ & $$ \\
$1$ & $e + \frac{3}{8} e^3 + \frac{65}{192} e^5$ & $-e + \frac{3}{8} e^3 + \frac{7}{192} e^5$ & $\frac{3}{2} e + \frac{27}{16} e^3 + \frac{261}{128} e^5$ & $- \frac{1}{2} e + \frac{1}{16} e^3 - \frac{5}{384} e^5$ \\
$2$ & $\frac{5}{4} e^2 + \frac{1}{6} e^4 + \frac{21}{64} e^6$ & $1 - \frac{7}{2} e^2 + \frac{29}{16} e^4 - \frac{53}{288} e^6$  & $\frac{9}{4} e^2 + \frac{7}{4} e^4 + \frac{141}{64} e^6$ & $1 - \frac{5}{2} e^2 + \frac{13}{16} e^4 - \frac{35}{288} e^6$ \\
$3$ & $\frac{13}{8} e^3 - \frac{25}{128} e^5$ & $ 3 e - \frac{69}{8} e^3 + \frac{369}{64} e^5$ & $\frac{53}{16} e^3 + \frac{393}{256} e^5$ & $\frac{7}{2} e - \frac{123}{16} e^3 + \frac{489}{128} e^5$ \\
$4$ & $\frac{103}{48} e^4 - \frac{129}{160} e^6$ & $\frac{13}{2} e^2 - \frac{55}{3} e^4 + \frac{239}{16} e^6$ & $\frac{77}{16} e^4 + \frac{129}{160} e^6$ & $\frac{17}{2} e^2 - \frac{115}{6} e^4 + \frac{601}{48} e^6$ \\
$5$ & $\frac{1097}{384} e^5$ & $\frac{295}{24} e^3 - \frac{13745}{384} e^5$ & $\frac{1773}{256} e^5$ & $\frac{845}{48} e^3 - \frac{32525}{768} e^5$ \\
$6$ & $\frac{1223}{320} e^6$ & $\frac{345}{16} e^4 - \frac{10569}{160} e^6$ & $\frac{3167}{320} e^6$ & $\frac{533}{16} e^4 - \frac{13827}{160} e^6$ \\
$7$ & $$ & $\frac{69251}{1920} e^5$ & $$ & $\frac{228347}{3840} e^5$ \\
$8$ & $$ & $\frac{42037}{720} e^6$ & $$ & $\frac{73369}{720} e^6$ \\ \hline
\end{tabular} 
\end{center}
\caption{Hansen coefficients up to $ e^6 $. \llabel{tabHansen} }  
$X_k^{\ell,-2} (e) = X_{-k}^{\ell,2} (e)$. The exact expression of these coefficients is given by expression (\ref{210910d}).
\end{table*}

The coefficients of the inertia tensor $\hat \TI (\sigma)$ (Eqs.\,(\ref{200117c})$-$(\ref{200117h})) are given in the frequency domain. 
In order to use them in the equations of motion (Sect.~\ref{pmp}), we need to return to the time domain using an inverse Fourier transform (Eq.\,(\ref{200116d})):
\be
\TI (t) = \sqpi \iinf \hat \TI (\sigma) \, \ep^{\ii \sigma t} \, d \sigma 
\ . \llabel{210910a}
\ee
In general, the perturbations introduced by the forcing tensor $\AT (t)$ are quasi-periodic (Eq.\,(\ref{220210f})). 
Then, only a discrete number of frequencies exist, and we can express $\TI (t)$ as a series:
\be
\TI (t) = \sum_k \hat \TI (\sigma_k) \, \ep^{\ii \sigma_k t} 
\ . \llabel{210910b}
\ee
In the frame $(\vp,\vq,\vs)$, the position of the star only depends on the orbital motion (Eq.\,(\ref{210804e})), and so the only forcing frequencies are the orbital mean motion, $n$, and its harmonics.
Thus, we can express $\AT (t)$ and $\TI (t)$ through the Hansen coefficients, $X_k^{\ell,m}$,
\be
\left( \frac{r_\osdot}{a} \right)^\ell \ep^{\ii m \up} = \sumk X_k^{\ell,m}(e) \, \ep^{\ii k M} 
\ , \llabel{210910c}
\ee
with
\be
\llabel{210910d}
\begin{split}
X_k^{\ell,m} (e) & = \frac{1}{2 \pi} \int_{-\pi}^\pi \left( \frac{r_\osdot}{a} \right)^\ell \ep^{\ii (m \up-k M)} \, d M \\ 
& = \frac{1}{\pi} \int_0^\pi \left( \frac{1-e^2}{1+e \cos \up} \right)^\ell \cos (m \up-k M) \, d M \ . 
\end{split}
\ee
In Table~\ref{tabHansen}, we provide the expression of the Hansen coefficients used in this work expanded up to $e^6$.

For instance, for the $\AT_{23}$ coefficient (Eq.\,(\ref{220210f})), we get from expression (\ref{210804e})
\be
\begin{split}
\AT_{23} &= - \frac{4 \pi R^4}{5 g}  \left( \frac{a}{r_\osdot} \right)^2 \xj  \xk 
= \frac{4 \pi R^4}{5 g}  \left( \frac{a}{r_\osdot} \right)^2  \sin \thetao \cos \thetao \sin^2 (\vpi+\up) \\
 & = \frac{\pi R^4}{5 g}  \sin \thetao \cos \thetao  \left( \frac{a}{r_\osdot} \right)^2 \left(2 -  \ep^{\ii 2 (\vpi+\up)} - \ep^{-\ii 2 (\vpi+\up)}  \right) \\
& = \frac{\pi R^4}{5 g} \sin \thetao \, \sumk \cT \left(2 X_k^{-2,0} -  \ep^{\ii 2 \vpi} X_k^{-2,2} - \ep^{-\ii 2 \vpi} X_k^{-2,-2}  \right)  \ep^{\ii k M}
\ , \llabel{211001e} 
\end{split}
\ee
where $\cT = \cos \thetao = \vk \cdot \vs $.

Similarly, for the $\AT_{13}$ coefficient (Eq.\,(\ref{220210f})), we get
\be
\AT_{13} =  \ii \frac{\pi R^4}{5 g} \sin \thetao \, \sumk  \left( \ep^{-\ii 2 \vpi} X_k^{-2,-2}  - \ep^{\ii 2 \vpi} X_k^{-2,2} \right) \ep^{\ii k M}
\ . \llabel{211001d} 
\ee
Then, for the $I_{23}$ coefficient we finally have (Eqs.\,(\ref{200117c}) and (\ref{210910b}))
\be
\llabel{211001f} 
\begin{split}
I_{23}  = \frac{\pi R^4}{10 g}  \sin \thetao \sumk & \Bigg\{ 
\hat p_2 (kn-\om) \, \bigg[ 2 \cT X_k^{-2,0}  - (1+\cT) \, \ep^{\ii 2 \vpi} X_k^{-2,2} \\ 
& \quad \quad  + (1-\cT) \, \ep^{-\ii 2 \vpi} X_k^{-2,-2}  \bigg] \\
& + \hat p_2 (kn+\om) \, \bigg[ 2 \cT X_k^{-2,0} + (1- \cT) \, \ep^{\ii 2 \vpi} X_k^{-2,2} \\ 
& \quad \quad - (1+\cT) \, \ep^{-\ii 2 \vpi} X_k^{-2,-2}  \bigg] \, \Bigg\} \, \ep^{\ii k M}  
\ . 
\end{split}
\ee
For the remaining coefficients of $ \TI (t)$, we get (Eqs.\,(\ref{200117b})$-$(\ref{200117h}))
\be
\llabel{221004w} 
\begin{split}
I_{13} = \ii \frac{\pi R^4}{10 g}  \sin \thetao \sumk & \Bigg\{  \hat p_2 (kn-\om) \, \bigg[ 2 \cT X_k^{-2,0} - (1+\cT) \, \ep^{\ii 2 \vpi} X_k^{-2,2}  \\
& \quad \quad +  (1-\cT) \, \ep^{-\ii 2 \vpi} X_k^{-2,-2} \bigg] \\
& - \hat p_2 (kn+\om) \, \bigg[ 2 \cT X_k^{-2,0} +  (1-\cT) \, \ep^{\ii 2 \vpi} X_k^{-2,2}  \\
& \quad \quad -  (1+\cT) \,  \ep^{-\ii 2 \vpi} X_k^{-2,-2} \bigg] \, \Bigg\} \, \ep^{\ii k M}  \ ,
\end{split}
\ee
\be
\llabel{221004x} 
\begin{split}
I_{12} = \ii \frac{\pi R^4}{20 g} \sumk & \Bigg\{  \hat p_2 (kn-2\om) \, \bigg[ 2 (1- \cT^2) \, X_k^{-2,0}  \\
&  + (1+\cT)^2 \, \ep^{\ii 2 \vpi} X_k^{-2,2}  +  (1-\cT)^2 \, \ep^{-\ii 2 \vpi} X_k^{-2,-2} \bigg] \\
& - \hat p_2 (kn+2\om) \, \bigg[ (1+\cT)^2 \, \ep^{-\ii 2 \vpi} X_k^{-2,-2}   \\
&  +  (1-\cT)^2 \, \ep^{\ii 2 \vpi} X_k^{-2,2} + 2 (1- \cT^2) \, X_k^{-2,0}  \bigg] \, \Bigg\} \, \ep^{\ii k M}  \ ,
\end{split}
\ee
\be
\llabel{221004w} 
\begin{split}
I_{33} = \frac{\pi R^4}{5 g} \sumk & \Bigg\{  \hat p_2 (kn) \, \bigg[ 2 \left(\cT^2 - \frac13\right) X_k^{-2,0}  \\
& + (1- \cT^2) \, \bigg( \ep^{\ii 2 \vpi} X_k^{-2,2}  +  \ep^{-\ii 2 \vpi} X_k^{-2,-2} \bigg) \bigg] \, \Bigg\} \, \ep^{\ii k M}  \ ,
\end{split}
\ee
\be
I_{22} = \Delta I - \frac{I_{33}}{2}
\quad \mathrm{and} \quad
I_{11} = - \Delta I - \frac{I_{33}}{2}
\llabel{221004v}  \ ,
\ee
with
\be
\llabel{221004z} 
\begin{split}
\Delta I = \frac{\pi R^4}{20 g} \sumk & \Bigg\{  \hat p_2 (kn-2\om) \, \bigg[ 2 (1- \cT^2) \, X_k^{-2,0}  \\
&  + (1+\cT)^2 \, \ep^{\ii 2 \vpi} X_k^{-2,2}  +  (1-\cT)^2 \, \ep^{-\ii 2 \vpi} X_k^{-2,-2} \bigg] \\
& + \hat p_2 (kn+2\om) \, \bigg[ (1+\cT)^2 \, \ep^{-\ii 2 \vpi} X_k^{-2,-2}   \\
&  +  (1-\cT)^2 \, \ep^{\ii 2 \vpi} X_k^{-2,2} + 2 (1- \cT^2) \, X_k^{-2,0}  \bigg] \, \Bigg\} \, \ep^{\ii k M}  \ .
\end{split}
\ee

The orbital and spin evolution of the planet under the action of thermal atmospheric tides is completely described by the set of equations (\ref{151028c}), (\ref{210805b}) and $(\ref{211001f})$-$(\ref{221004z})$.

\section{Secular dynamics}
\llabel{secdyn}

In general, the thermal atmospheric tides slowly modify the spin and the orbit of the planet, in a timescale much longer than the orbital and precession periods of the system.
Therefore, we can average the equations of motion (section~\ref{pmp}) over the mean anomaly and the argument of the pericenter, and obtain the equations of motion for the secular evolution of the system.

\subsection{Averaging process}
\llabel{singav}

Following \citet{Correia_Valente_2022},
to average the equations of motion, we first expand them in Hansen coefficients (Eq.\,(\ref{210910c})), similarly to what we have done with the inertia tensor (Sect~\ref{hce}).
For instance, for the last term in the $\vs$ component of the tidal torque we have (Eq.\,(\ref{151028e}))
\be
\begin{split}
 - \frac{3 \Gc \Ms}{r^3} & I_{23} \xii \xk 
 = \frac{3 \Gc \Ms}{2 r^3} I_{23}  \sin \thetao \sin (2 \vpi+2\up) \\
& = \frac{3 \Gc \Ms}{4 \ii a^3} I_{23}  \sin \thetao \, \sumkl \left( \ep^{\ii 2 \vpi} X_\kl^{-3,2} - \ep^{-\ii 2 \vpi} X_\kl^{-3,-2}  \right) \ep^{\ii \kl M}
\ . \llabel{211001c}
\end{split}
\ee
Then, we replace the expression of $I_{23}$ also expanded in Hansen coefficients (Eq.\,(\ref{211001f})), and average over the mean anomaly, $M$, which is equivalent to retaining only the terms with $\kl = -k$, and over the argument of the pericenter
\be
 \llabel{211006a}
\begin{split}
 \left\langle - \frac{3 \Gc \Ms}{r^3} I_{23} \xii \xk \right\rangle_{M,\vpi}   & = \\
 \frac{3 \pi \Gc \Ms R^4}{40 \ii a^3 g}  \sin^2 \thetao \sumk & \Bigg\{ \hat p_2 (kn-\om) \, \bigg[ (1+\cT) \, X_k^{-2,2} X_{-k}^{-3,-2}  \\ 
& \quad \quad + (1-\cT) \, X_k^{-2,-2} X_{-k}^{-3,2} \bigg]  \\
& - \hat p_2 (kn+\om) \, \bigg[ (1- \cT) \, X_k^{-2,2} X_{-k}^{-3,-2}  \\
& \quad \quad + (1+\cT) \,  X_k^{-2,-2} X_{-k}^{-3,2} \bigg] \Bigg\} \ .
\end{split}
\ee
Finally, we decompose the surface pressure variations $\hat p_2 (k n \pm \om)$ in its real and imaginary parts (Eq.\,(\ref{210924d})), make use of their parity properties, and use the simplification $X_{-k}^{\ell,m} = X_k^{\ell,-m}$ to write
\be
 \llabel{211103f}
\begin{split}
& \left\langle - \frac{3 \Gc \Ms}{r^3} I_{23} \xii \xk \right\rangle_{M,\vpi}  = 
 \frac{3 \pi \Gc \Ms R^4}{20 a^3 g}  \sin^2 \thetao \, \, \times \\ 
 &  \sumk \bu \, \bigg[ (1+\cT) \, X_k^{-2,2} X_k^{-3,2} + (1- \cT) \, X_k^{-2,-2} X_k^{-3,-2}  \bigg] \ .
\end{split}
\ee
This last arrangement is very useful, because we are able to combine terms in $b (\om \pm kn)$ in a single term $b (\om-kn)$, which considerably simplifies the expression of the equations of motion.
In this particular case, we also note that there is no longer the contribution from $\au$.

Another simplification is that we do not need to follow the evolution of the position vector anymore (Eq.\,(\ref{151028c})).
Indeed, as we average over the mean anomaly, $M$, the position of the planet in the orbit is no longer defined, and as we average over $\vpi$, the position of the pericenter is also no longer defined.
Therefore, in the secular case, the equations of motion can be given by the evolution of the orbital angular momentum (Eq.\,(\ref{210804a})) together with the evolution of the eccentricity.
Alternatively, we prefer to use the evolution of the orbital energy, 
\be
\Eo = - \frac{\Gc \Ms \ms}{2 a} 
\ , \label{210804a2}
\ee
because it can be directly obtained from the power of the tidal force (Eq.\,(\ref{170911d})) as
\be
\Po = \dot \vr_\star \cdot \vF
\ , \llabel{211110b}
\ee
and thus provides a simpler expression, from which we can later easily derive the evolution of the eccentricity.

\subsection{Tidal torque and power}
\llabel{sec_ttp}

The orbital and spin evolution of the planet are completely described by the evolution of the angular momentum vectors, $\vG$ and $\vL$, given by the torque (Eqs.\,(\ref{150626a}) and (\ref{210805b})), and by the evolution of the orbital energy, $\Eo$, given by the power (Eqs.\,(\ref{211110b})).

The secular torque (Eq.\,(\ref{151028e})) can be written as 
\be
\big\langle \vT \big\rangle_{M,\vpi} =   T_p \, \vp +  T_q \, \vq +  T_s \, \vs 
\ . \llabel{220405a}
\ee
For zero obliquity ($\thetao=0$), the vectors $\vp$ and $\vq$ of the basis are not defined (Eq.\,(\ref{221007a})).
There are no singularities in this problem because both $ T_p,  T_q \propto \sin \thetao$; however, to avoid numerical issues, it is easier to express the torque simply using the unit vectors of the angular momentum quantities, $\vk$ and $\vs$, as
\be
\big\langle \vT \big\rangle_{M,\vpi}   = \overline T_1 \, \vk + \overline T_2 \, \vs + \overline T_3 \, \vk \times \vs 
\ , \llabel{211110a}
\ee
where
\be
\overline T_1 =  \frac{ T_q}{\sin \thetao}  \ , \quad
\overline T_2 =  T_s - \cos \thetao \frac{ T_q}{\sin \thetao} \ , \quad
\overline T_3 = \frac{ T_p}{\sin \thetao} 
\ ,  \llabel{211017e}
\ee
with
\be
\llabel{211110t1}
\begin{split}
\overline T_1 =  
- \frac{3 \At}{32} \sumk & \Bigg\{ 3 \, \bz  \bigg[
\left(1-\cT^2\right) \left( \Xtdp - \Xtdm \right) \bigg]  \\
&    +  2 \, \bu  \bigg[   
   \left(1-\cT\right)^2 \left(2+\cT\right) \Xtdm \\
& + 4 \cT^3 \Xtz 
 - \left(1+\cT\right)^2 \left(2-\cT\right) \Xtdp 
 \bigg]  \\
&   +  \bd  \bigg[   
   4 \cT \left(1-\cT^2\right) \Xtz \\
&  + \left(1-\cT\right)^3 \Xtdm 
 - \left(1+\cT\right)^3 \Xtdp 
 \bigg] \Bigg\} \ ,
\end{split}
\ee
\be
\llabel{211110t2}
\begin{split}
\overline T_2 = 
- \frac{3 \At}{32} \sumk & \Bigg\{  3 \, \bz \bigg[
\cT \left(1-\cT^2\right) \left(\Xtdm  - \Xtdp \right) 
\bigg] \\
&  - 2 \, \bu  \bigg[   
  \left(1-\cT\right)^2 \left(1+2 \cT\right) \Xtdm \\
&  + 4 \cT^2 \Xtz 
 + \left(1+\cT\right)^2 \left(1-2 \cT\right) \Xtdp 
 \bigg]  \\
&  -  \bd  \bigg[   
  \left(1-\cT\right)^3 \Xtdm \\
& + 4 \left(1-\cT^2\right) \Xtz 
 + \left(1+\cT\right)^3 \Xtdp 
 \bigg] \Bigg\} \ ,
\end{split}
\ee
\be
\llabel{211110t3}
\begin{split}
\overline T_3 = 
\frac{3 \At}{32} \sumk & \Bigg\{ \cT \, \az  \bigg[
  - 4 \left(1-3 \cT^2\right) \Xtz \\ &
 - 3 \left(1-\cT^2\right) \left(\Xtdm  + \Xtdp \right)
 \bigg]  \\
&  + 2 \, \au  \bigg[   
 - \left(1-\cT\right)^2 \left(1+2 \cT\right) \Xtdm \\ &
 + 4 \cT \left(1-2 \cT^2\right) \Xtz 
 + \left(1+\cT\right)^2 \left(1-2 \cT\right) \Xtdp 
 \bigg]  \\
&  - \ad  \bigg[   
   4 \cT \left(1-\cT^2\right) \Xtz \\ &
 + \left(1-\cT\right)^3 \Xtdm 
 - \left(1+\cT\right)^3 \Xtdp 
 \bigg] \Bigg\} \ ,
\end{split}
\ee
and 
\be
\At = \frac{4 \pi \Gc \Ms R^4}{5 a^3 g} = \frac{3 \Ms}{5 \bar \rho} \left( \frac{R}{a} \right)^3 \ ,
\ee
where $\bar \rho = 3 g / (4 \pi \Gc R)$ is the mean density of the planet.
These expressions were obtained using the algebraic manipulators \citet{Maxima_2022} and TRIP \citep{Gastineau_Laskar_2011}. 

Finally, for the secular power (Eq.\,(\ref{211110b})) we get
\be
\llabel{211110t4}
\begin{split}
\big\langle \Po  \big\rangle_{M,\vpi} =
- \dfrac{\At}{64} \sumk & k n \Bigg\{ \bz  \bigg[
   4 \left(1-3 \cT^2\right)^2 \Xtz \\ &
 + 9 \left(1-\cT^2\right)^2 \left(\Xtdm  + \Xtdp \right)
 \bigg]  \\
&  - 12 \, \bu \left(1-\cT^2\right) \bigg[   
   4 \cT^2  \Xtz \\ &
 +  \left(1-\cT\right)^2 \Xtdm 
 + \left(1+\cT\right)^2 \Xtdp  
 \bigg]  \\
&  - 3 \,\bd  \bigg[   
   4 \left(1-\cT^2\right)^2 \Xtz \\ &
 + \left(1-\cT\right)^4 \Xtdm  
 + \left(1+\cT\right)^4 \Xtdp 
 \bigg] \Bigg\} \ .
\end{split}
\ee

\subsection{Orbital and spin evolution}
\llabel{oase2}

The set of equations (\ref{211110a}) and (\ref{211110t4}) allows us to track the secular evolution of the system using the angular momentum vectors and the orbital energy.
For a more intuitive description of the orbital and spin evolution, we can relate these quantities to the orbital elements and the rotation of the planet.
The semimajor axis is directly given from the orbital energy (Eq.\,(\ref{210804a2}))
\be
a = - \frac{\beta \mu}{2 \Eo}
\ , \llabel{211110f}
\ee
the eccentricity from the orbital angular momentum (Eq.\,(\ref{210804a}))
\be
e  = \sqrt{1 - \frac{(\vG \cdot \vk)^2}{\beta^2 \mu a}}
= \sqrt{1 - \frac{\vG \cdot \vG}{\beta^2 \mu a}}
\ , \llabel{211110g}
\ee
and the rotation rate from the rotational angular momentum (Eq.\,(\ref{151019b})),
\be
\om = \frac{\vL \cdot \vs}{C} 
= \frac{\sqrt{\vL \cdot \vL}}{C}
\ . \llabel{211027c}
\ee
The angle between the orbital and equatorial planes (also known as obliquity or inclination), can be obtained from both angular momentum vectors as (Fig.~\ref{frames})
\be
\cos \thetao = \vk \cdot \vs = \frac{\vG \cdot \vL}{\sqrt{(\vG \cdot \vG) (\vL \cdot \vL)}} 
\ . \llabel{211027z}
\ee

For a better comparison with previous studies, we can also directly obtain the evolution of all these quantities.
The semimajor axis evolution is given from expression (\ref{211110t4}),
\be
\dot a = \frac{2 a^2}{\beta \mu} \, \Po 
\ , \llabel{211110c}
\ee
while the eccentricity evolution can be computed from expressions (\ref{211110a}), (\ref{211110g}) and (\ref{211110c}) as
\be
\dot e =  \frac{1-e^2}{2 a e} \, \dot a  - \frac{\vG \cdot \vT}{\beta^2 \mu a e} 
=  \frac{\sq}{\beta n a^2 e} \left( \frac{\sq}{n} \, \Po - \overline T_1 - \overline T_2 \, \cT \right)
\ , \llabel{211110d}
\ee
that is, using expressions (\ref{211110t1}), (\ref{211110t2}), and (\ref{211110t4}), 
\be
\llabel{211110e}
\begin{split}
\dot e  = - \dfrac{\At}{64} \, \frac{\sq}{\beta n a^2 e} & \sumk  \Bigg\{ \bz  \bigg[ 
   4 \left(1-3\cT^2\right)^2 \Xtz  k \sq \\&
 + 9 \left(1-\cT^2\right)^2 \Xtdm  \left(2+ k \sq\right)  \\ &
 - 9 \left(1-\cT^2\right)^2 \Xtdp  \left(2- k \sq\right)  \bigg]  \\
 &  - 12 \, \bu \left(1-\cT^2\right) \bigg[   
    4 \cT^2 \Xtz  k \sq  \\&
 + \left(1-\cT\right)^2 \Xtdm  \left(2+k \sq\right)  \\ &
 -  \left(1+\cT\right)^2 \Xtdp  \left(2-k \sq\right)  \bigg]  \\
  & - 3 \, \bd  \bigg[   
 4 \left(1-\cT^2\right)^2 \Xtz  k \sq   \\ &
 + \left(1-\cT\right)^4 \Xtdm  \left(2+k \sq\right)   \\ &
 - \left(1+\cT\right)^4 \Xtdp  \left(2-k \sq\right)  \bigg] \Bigg\} \ .
\end{split}
\ee
For the rotation rate evolution, we have the following from Eqs.~(\ref{210805b}) and (\ref{211027c}):
\be
\dot \om = - \frac{\vT \cdot \vs}{C} 
= - \frac{\overline T_1 \, \cT + \overline T_2}{C} 
\ , \llabel{211110h}
\ee
that is, using expressions (\ref{211110t1}) and (\ref{211110t2})
\be
\llabel{211110t5}
\begin{split}
\dot \om = - \frac{3 \At}{32 C} \sum_{k=-\infty}^{+\infty} & \Bigg\{ 2 \, \bu \left( 1 - \cT^2 \right) \bigg[ 4 \cT^2 \Xtz    \\
   & + \left( 1 + \cT \right)^2 \Xtdp  + \left( 1 -\cT \right)^2 \Xtdm   \bigg] \\
   & + \bd \bigg[ 4 \left( 1 - \cT^2 \right)^2 \Xtz   \\
   & + \left( 1 + \cT  \right)^4 \Xtdp  + \left( 1 - \cT  \right)^4 \Xtdm   \bigg] \Bigg\}   \ .
\end{split}
\ee
The obliquity (or inclination) evolution is given from expressions (\ref{150626a}), (\ref{210805b}) and (\ref{211027z}),
\be
\llabel{211110i}
\begin{split}
\dot \thetao & = \frac{\vT \cdot \vk - \vT \cdot \vs \, \cos \thetao}{\sin \thetao \sqrt{\vL \cdot \vL}} - \frac{\vT \cdot \vs - \vT \cdot \vk \, \cos \thetao}{\sin \thetao \sqrt{\vG \cdot \vG}}  \\ &
 = \left( \frac{\overline T_1}{C \om} - \frac{\overline T_2}{\beta \sqrt{\mu a (1-e^2)}} \right)  \sin \thetao
 \approx \frac{\overline T_1}{C \om} \sin \thetao
\ . 
\end{split}
\ee
In general, for a planet around a star, we have $| \vL | \ll | \vG |$, and so the evolution is dominated by the first term.

Finally, we can also obtain the evolution of the precession angles, that is, the angular velocity of the longitude of the node, $ \dot \Omega$, and the precession speed of the spin axis, $\dot \psi$.
The line of nodes is aligned with the vector $\vp$ (Fig.~\ref{frames}) and thus
\be
\dot \Omega = \frac{ \dot \vG}{|\vG|} \cdot \vp =\frac{\vT \cdot (\vk \times \vs)}{|\vG| \sin \thetao} 
=  \frac{\overline T_3 \sin \thetao}{\beta \sqrt{\mu a (1-e^2)}}
\ , \llabel{211110j}
\ee
\be
\dot \psi  = \frac{ \dot \vL}{|\vL|} \cdot \vp = - \frac{\vT \cdot (\vk \times \vs)}{|\vL| \sin \thetao} 
= - \frac{\overline T_3 \sin \thetao}{C \om}
\ . \llabel{211110k}
\ee

\subsection{Energy balance}

The average total energy transferred to the planet due to thermal atmospheric tides is given by
\be
\Pt = \big\langle \Po + \dot \Er \big\rangle_{M,\vpi} 
\ , \llabel{211029d}
\ee
where $\Eo$ is the orbital energy (Eq.\,(\ref{210804a2})), 
\be 
\Er = \frac{\vw \cdot \vL}{2}
\llabel{211029er}
\ee
is the rotational energy, and
\be
\dot \Er = C \om \, \dot \om
\ , \llabel{211029fr}
\ee
where $\dot \om$ is given by expression (\ref{211110h}).

The total energy is then obtained by combining expressions (\ref{211110t4}) and (\ref{211110t5}) as
\be
\llabel{211110l}
\begin{split}
\Pt = - \dfrac{\At}{64}  \sumk & \Bigg\{  k n \,\bz  \bigg[
   4 \left(1-3 \cT^2\right)^2 \Xtz \\& 
 + 9 \left(1-\cT^2\right)^2 \left(\Xtdm  + \Xtdp \right)
 \bigg]  \\
&  + 12 \, (\om-k n) \, \bu \left(1-\cT^2\right) \bigg[   
   4 \cT^2  \Xtz \\ &
 +  \left(1-\cT\right)^2 \Xtdm 
 + \left(1+\cT\right)^2 \Xtdp  
 \bigg]  \\
&  + 3 \, (2 \om -k n) \, \bd  \bigg[   
   4 \left(1-\cT^2\right)^2 \Xtz  \\ &
 + \left(1-\cT\right)^4 \Xtdm  
 + \left(1+\cT\right)^4 \Xtdp 
 \bigg] \Bigg\} \ .
\end{split}
\ee

We note that regardless of the orbital and spin parameters, the total energy transferred is always positive since $b(\sigma)$ is an odd function and $b(|\sigma|) < 0$ (Eq.\,(\ref{210929z})).

\section{Expansion up to $e^2$}

\llabel{expe2}

Most Earth-mass planets are observed in multiplanet systems, whose eccentricities are relatively small for stability reasons.
Therefore, to simplify the equations of motion, we can truncate the series expansion in Hansen coefficients, and retain only terms in $e^2$ or smaller.

\subsection{Tidal torque and power}

For the average tidal torque (Eq.\,(\ref{211110a})), we have
\be
\llabel{221105t1}
\begin{split}
\overline T_1 = - \At & \Bigg\{
  \bZwpTn \frac{189}{32} \left(1-\cT^2\right) e^2 \\ & 
 + \bZwpDn \frac{9}{16} \left(1-\cT^2\right) \left(1-6 e^2\right) \\ &
 + \bZwpUn \frac{9}{32} \left(1-\cT^2\right) e^2 \\ &
 + \bUwpTn \frac{63}{32} \left(1-\cT\right)^2 \left(2+\cT\right) e^2 \\ &
 + \bUwpDn \frac{3}{16} \left(1-\cT\right)^2 \left(2+\cT\right) \left(1-6 e^2\right) \\ &
 + \bUwpUn \frac{3}{32} \left(2-3 \cT+13 \cT^3\right) e^2 \\ &
  + \bUwpZn \frac{3}{4} \cT^3 \left(1+2 e^2\right) \\ &
 - \bUwmUn \frac{3}{32} \left(2+3 \cT-13 \cT^3\right) e^2 \\ &
 - \bUwmDn \frac{3}{16} \left(1+\cT\right)^2 \left(2-\cT\right) \left(1-6 e^2\right) \\ &
 - \bUwmTn \frac{63}{32} \left(1+\cT\right)^2 \left(2-\cT\right) e^2 \\ &
 + \bDwpTn \frac{63}{64} \left(1-\cT\right)^3 e^2 \\ &
 + \bDwpDn \frac{3}{32} \left(1-\cT\right)^3 \left(1-6 e^2\right) \\ &
 + \bDwpUn \frac{3}{64} \left(1-\cT\right) \left(1+10 \cT+13 \cT^2\right) e^2 \\ &
 + \bDwpZn \frac{3}{8} \cT \left(1-\cT^2\right) \left(1+2 e^2\right) \\ &
 - \bDwmUn \frac{3}{64} \left(1+\cT\right) \left(1-10 \cT+13 \cT^2\right) e^2 \\ &
 - \bDwmDn \frac{3}{32} \left(1+\cT\right)^3 \left(1-6 e^2\right) \\ &
 - \bDwmTn \frac{63}{64} \left(1+\cT\right)^3 e^2
 \Bigg\} \ ,
\end{split}
\ee
\be
\llabel{221105t2}
\begin{split}
\overline T_2 = \At & \Bigg\{ 
  \bZwpTn \frac{189}{32} \cT \left(1-\cT^2\right) e^2 \\ &
 + \bZwpDn \frac{9}{16} \cT \left(1-\cT^2\right) \left(1-6 e^2\right) \\ &
 + \bZwpUn \frac{9}{32} \cT \left(1-\cT^2\right) e^2 \\ &
 + \bUwpTn \frac{63}{32} \left(1-\cT\right)^2 \left(1+2 \cT\right) e^2 \\ &
 + \bUwpDn \frac{3}{16} \left(1-\cT\right)^2 \left(1+2 \cT\right) \left(1-6 e^2\right) \\ &
 + \bUwpUn \frac{3}{32} \left(1+9 \cT^2+2 \cT^3\right) e^2 \\ &
 + \bUwpZn \frac{3}{4} \cT^2 \left(1+2 e^2\right) \\ &
 + \bUwmUn \frac{3}{32} \left(1+9 \cT^2-2 \cT^3\right) e^2 \\ &
 + \bUwmDn \frac{3}{16} \left(1+\cT\right)^2 \left(1-2 \cT\right) \left(1-6 e^2\right) \\ &
 + \bUwmTn \frac{63}{32} \left(1+\cT\right)^2 \left(1-2 \cT\right) e^2 \\ &
 + \bDwpTn \frac{63}{64} \left(1-\cT\right)^3 e^2 \\ &
 + \bDwpDn \frac{3}{32} \left(1-\cT\right)^3 \left(1-6 e^2\right) \\ &
 + \bDwpUn \frac{3}{64} \left(1-\cT\right) \left(13+10 \cT+\cT^2\right) e^2 \\ &
 + \bDwpZn \frac{3}{8} \left(1-\cT^2\right) \left(1+2 e^2\right) \\ &
 + \bDwmUn \frac{3}{64} \left(1+\cT\right) \left(13-10 \cT+\cT^2\right) e^2 \\ &
 + \bDwmDn \frac{3}{32} \left(1+\cT\right)^3 \left(1-6 e^2\right) \\ &
 + \bDwmTn \frac{63}{64} \left(1+\cT\right)^3 e^2
 \Bigg\} \ ,
\end{split}
\ee
\be
\llabel{221105t3}
\begin{split}
\overline T_3 = - \At & \Bigg\{  
  \aZwpTn \frac{189}{32} \cT \left(1-\cT^2\right) e^2 \\ &
 + \aZwpDn \frac{9}{16} \cT \left(1-\cT^2\right) \left(1-6 e^2\right) \\ &
 + \aZwpUn \frac{9}{32} \cT \left(5-13 \cT^2\right) e^2 \\ &
 + \aZwpZn \frac{3}{8} \cT \left(1-3 \cT^2\right) \left(1+2 e^2\right) \\ &
 + \aUwpTn \frac{63}{32} \left(1-\cT\right)^2 \left(1+2 \cT\right) e^2 \\ &
 + \aUwpDn \frac{3}{16} \left(1-\cT\right)^2 \left(1+2 \cT\right) \left(1-6 e^2\right) \\ &
 + \aUwpUn \frac{3}{32} \left(1-12 \cT-3 \cT^2+26 \cT^3\right) e^2 \\ &
  - \aUwpZn \frac{3}{4} \cT \left(1-2 \cT^2\right) \left(1+2 e^2\right) \\ &
 - \aUwmUn \frac{3}{32} \left(1+12 \cT-3 \cT^2-26 \cT^3\right) e^2 \\ &
 - \aUwmDn \frac{3}{16} \left(1+\cT\right)^2 \left(1-2 \cT\right) \left(1-6 e^2\right) \\ &
 - \aUwmTn \frac{63}{32} \left(1+\cT\right)^2 \left(1-2 \cT\right) e^2 \\ &
 + \aDwpTn \frac{63}{64} \left(1-\cT\right)^3 e^2 \\ &
 + \aDwpDn \frac{3}{32} \left(1-\cT\right)^3 \left(1-6 e^2\right) \\ &
 + \aDwpUn \frac{3}{64} \left(1-\cT\right) \left(1+10 \cT+13 \cT^2\right) e^2 \\ &
 + \aDwpZn \frac{3}{8} \cT \left(1-\cT^2\right) \left(1+2 e^2\right) \\ &
 - \aDwmUn \frac{3}{64} \left(1+\cT\right) \left(1-10 \cT+13 \cT^2\right) e^2 \\ &
 - \aDwmDn \frac{3}{32} \left(1+\cT\right)^3 \left(1-6 e^2\right) \\ &
 - \aDwmTn \frac{63}{64} \left(1+\cT\right)^3 e^2 
 \Bigg\} \ ,
\end{split}
\ee
and for the average power (Eq.\,(\ref{211110t4})) 
\be
\llabel{221105t4}
\begin{split}
\big\langle \Po  \big\rangle_{M,\vpi} = - n \At & \Bigg\{
  \bZwpTn \frac{567}{64} \left(1-\cT^2\right)^2 e^2 \\ &
 + \bZwpDn \frac{9}{16} \left(1-\cT^2\right)^2 \left(1-6 e^2\right) \\ &
 + \bZwpUn \frac{3}{64} \left(7-30 \cT^2+39 \cT^4\right) e^2 \\ &
 + \bUwpTn \frac{189}{32} \left(1-\cT^2\right) \left(1-\cT\right)^2 e^2 \\ &
 + \bUwpDn \frac{3}{8} \left(1-\cT^2\right) \left(1-\cT\right)^2 \left(1-6 e^2\right) \\ &
 + \bUwpUn \frac{3}{32} \left(1-\cT^2\right) \left(1-2 \cT+13 \cT^2\right) e^2 \\ &
 - \bUwmUn \frac{3}{32} \left(1-\cT^2\right) \left(1+2 \cT+13 \cT^2\right) e^2 \\ &
 - \bUwmDn \frac{3}{8} \left(1-\cT^2\right) \left(1+\cT\right)^2 \left(1-6 e^2\right) \\ &
 - \bUwmTn \frac{189}{32} \left(1-\cT^2\right) \left(1+\cT\right)^2 e^2 \\ &
 + \bDwpTn \frac{189}{128} \left(1-\cT\right)^4 e^2 \\ &
 + \bDwpDn \frac{3}{32} \left(1-\cT\right)^4 \left(1-6 e^2\right) \\ &
 + \bDwpUn \frac{3}{128} \left(1-\cT\right)^2 \left(13+22 \cT+13 \cT^2\right) e^2 \\ &
 - \bDwmUn \frac{3}{128} \left(1+\cT\right)^2 \left(13-22 \cT+13 \cT^2\right) e^2 \\ &
 - \bDwmDn \frac{3}{32} \left(1+\cT\right)^4 \left(1-6 e^2\right) \\ &
 - \bDwmTn \frac{189}{128} \left(1+\cT\right)^4 e^2
 \Bigg\} \ .
\end{split}
\ee

\subsection{Orbital and spin evolution}

The semimajor axis evolution can be directly obtained replacing Eq.\,(\ref{221105t4}) in Eq.\,(\ref{211110c}).
For the eccentricity, we have (Eq.\,(\ref{211110e}))
\be
\llabel{221105te}
\begin{split}
\dot e  =  - \dfrac{\At e}{\beta n a^2} &  \Bigg\{ 
 \bZwpTn \frac{189}{64} \left(1-\cT^2\right)^2 \\ &
 - \bZwpDn \frac{9}{32} \left(1-\cT^2\right)^2 \\ &
 + \bZwpUn \frac{3}{64} \left(1-18 \cT^2+33 \cT^4\right) \\ &
 + \bUwpTn \frac{63}{32} \left(1-\cT^2\right) \left(1-\cT\right)^2 \\ &
 - \bUwpDn \frac{3}{16} \left(1-\cT^2\right) \left(1-\cT\right)^2 \\ &
 - \bUwpUn \frac{3}{32} \left(1-\cT^2\right) \left(1-2 \cT-11 \cT^2\right) \\ &
 + \bUwmUn \frac{3}{32} \left(1-\cT^2\right) \left(1+2 \cT-11 \cT^2\right) \\ &
 + \bUwmDn \frac{3}{16} \left(1-\cT^2\right) \left(1+\cT\right)^2 \\ &
 - \bUwmTn \frac{63}{32} \left(1-\cT^2\right) \left(1+\cT\right)^2 \\ &
 + \bDwpTn \frac{63}{128} \left(1-\cT\right)^4 \\ &
 - \bDwpDn \frac{3}{64} \left(1-\cT\right)^4 \\ &
 + \bDwpUn \frac{3}{128} \left(1-\cT\right)^2 \left(11+26 \cT+11 \cT^2\right) \\ &
 - \bDwmUn \frac{3}{128} \left(1+\cT\right)^2 \left(11-26 \cT+11 \cT^2\right) \\ &
 + \bDwmDn \frac{3}{64} \left(1+\cT\right)^4 \\ &
 - \bDwmTn \frac{63}{128} \left(1+\cT\right)^4 
 \Bigg\} \ .
\end{split}
\ee
The evolution of the obliquity is directly obtained by replacing Eq.\,(\ref{221105t1}) in Eq.\,(\ref{211110i}), and for the rotation rate (Eq.\,(\ref{211110t5})),
\be
\llabel{221105tw}
\begin{split}
\dot \om =
- \frac{\At}{C} & \Bigg\{  
  \bUwpTn \frac{63}{32} \left(1-\cT^2\right) \left(1-\cT\right)^2 e^2 \\ &
 + \bUwpDn \frac{3}{16} \left(1-\cT^2\right) \left(1-\cT\right)^2 \left(1-6 e^2\right) \\ &
 + \bUwpUn \frac{3}{32} \left(1-\cT^2\right) \left(1-2 \cT+13 \cT^2\right) e^2 \\ &
 + \bUwpZn \frac{3}{4} \cT^2 \left(1-\cT^2\right) \left(1+2 e^2\right) \\ &
 + \bUwmUn \frac{3}{32} \left(1-\cT^2\right) \left(1+2 \cT+13 \cT^2\right) e^2 \\ &
 + \bUwmDn \frac{3}{16} \left(1-\cT^2\right) \left(1+\cT\right)^2 \left(1-6 e^2\right) \\ &
 + \bUwmTn \frac{63}{32} \left(1-\cT^2\right) \left(1+\cT\right)^2 e^2 \\ &
 + \bDwpTn \frac{63}{64} \left(1-\cT\right)^4 e^2 \\ &
 + \bDwpDn \frac{3}{32} \left(1-\cT\right)^4 \left(1-6 e^2\right) \\ &
 + \bDwpUn \frac{3}{64} \left(1-\cT\right)^2 \left(13+22 \cT+13 \cT^2\right) e^2 \\ &
 + \bDwpZn \frac{3}{8} \left(1-\cT^2\right)^2 \left(1+2 e^2\right) \\ &
 + \bDwmUn \frac{3}{64} \left(1+\cT\right)^2 \left(13-22 \cT+13 \cT^2\right) e^2 \\ &
 + \bDwmDn \frac{3}{32} \left(1+\cT\right)^4 \left(1-6 e^2\right) \\ &
 + \bDwmTn \frac{63}{64} \left(1+\cT\right)^4 e^2
 \Bigg\} \ .
\end{split}
\ee
This last expression for $e=0$ is equivalent to Eq.\,(11) in \citet {Dobrovolskis_1980} and Eq.\,(7) in \citet{Correia_Laskar_2003JGR}.
It should also match Eq.\,(19) in \citet{Cunha_etal_2015}, but they performed a development of the atmospheric tidal potential in $r_\osdot^{-5}$, while here the potential only depends on $r_\osdot^{-3}$ (Eq.\,(\ref{220210d})), and the extra factor in $r_\osdot^{-2}$ is included in the forcing function (Eq.\,(\ref{220210f})).

\subsection{Energy balance}

The total energy is obtained from expression (\ref{211110l}) as
\be
\llabel{221105t5}
\begin{split}
\Pt = - \At & \Bigg\{ 
  \left(3n\right) \bZwpTn \frac{189}{64} \left(1-\cT^2\right)^2 e^2 \\ &
 + \left(2n\right) \bZwpDn \frac{9}{32} \left(1-\cT^2\right)^2 \left(1-6 e^2\right) \\ &
 + \left(n\right) \bZwpUn \frac{3}{64} \left(7-30 \cT^2+39 \cT^4\right) e^2 \\ &
 + \left(\om+3n\right) \bUwpTn \frac{63}{32} \left(1-\cT^2\right) \left(1-\cT\right)^2 e^2 \\ &
 + \left(\om+2n\right) \bUwpDn \frac{3}{16} \left(1-\cT^2\right) \left(1-\cT\right)^2 \left(1-6 e^2\right) \\ &
 + \left(\om+n\right) \bUwpUn \frac{3}{32} \left(1-\cT^2\right) \left(1-2 \cT+13 \cT^2\right) e^2 \\ &
 + \left(\om\right) \bUwpZn \frac{3}{4} \cT^2 \left(1-\cT^2\right) \left(1+2 e^2\right) \\ &
 + \left(\om-n\right) \bUwmUn \frac{3}{32} \left(1-\cT^2\right) \left(1+2 \cT+13 \cT^2\right) e^2 \\ &
 + \left(\om-2n\right) \bUwmDn \frac{3}{16} \left(1-\cT^2\right) \left(1+\cT\right)^2 \left(1-6 e^2\right) \\ &
 + \left(\om-3n\right) \bUwmTn \frac{63}{32} \left(1-\cT^2\right) \left(1+\cT\right)^2 e^2 \\ &
 + \left(2\om+3n\right) \bDwpTn \frac{63}{128} \left(1-\cT\right)^4 e^2 \\ &
 + \left(2\om+2n\right) \bDwpDn \frac{3}{64} \left(1-\cT\right)^4 \left(1-6 e^2\right) \\ &
 + \left(2\om+n\right) \bDwpUn \frac{3}{128} \left(1-\cT\right)^2 \left(13+22 \cT+13 \cT^2\right) e^2 \\ &
 + \left(2\om\right) \bDwpZn \frac{3}{16} \left(1-\cT^2\right)^2 \left(1+2 e^2\right) \\ &
 + \left(2\om-n\right) \bDwmUn \frac{3}{128} \left(1+\cT\right)^2 \left(13-22 \cT+13 \cT^2\right) e^2 \\ &
 + \left(2\om-2n\right) \bDwmDn \frac{3}{64} \left(1+\cT\right)^4 \left(1-6 e^2\right) \\ &
 + \left(2\om-3n\right) \bDwmTn \frac{63}{128} \left(1+\cT\right)^4 e^2
\Bigg\} \ .
\end{split}
\ee

%%%%%%%%%%%%%%%%%%%%%%%

\section{Planar case}

\llabel{planarcase}

The final outcome of tidal evolution is the alignment of the spin axis with the normal to the orbit \citep[see][]{Correia_etal_2003}.
Therefore, to simplify the equations of motion, many works assume that this alignment is always present, that is, the motion is planar ($\thetao = 0$).
In this case, we have $\cT=1$ and $\vs = \vk$ (Fig.~\ref{frames}).

\subsection{Tidal torque and power}

Using the simplification $\vs = \vk$ yields (Eqs.\,(\ref{220405a}) and (\ref{211110a}))
\be
\big\langle \vT \big\rangle_{M,\vpi} = T_s \, \vk   = \Big(\overline T_1 + \overline T_2 \Big) \, \vk
\ , \llabel{211028h}
\ee
for the average tidal torque, with
\be
T_s =  \dfrac{3 \At}{2}  \sumk \bd \, \Xtdp  
\ . \llabel{211028j}
\ee
For the average power, we get the following from Eq.\,(\ref{211110t4}) with $x=1$:
\be
\llabel{211103c}
\begin{split}
\big\langle \Po  \big\rangle_{M,\vpi}  
= - \frac{\At}{4}  \sumk & kn \bigg[  \bz  \Xtz  
   - 3 \, \bd \Xtdp  \bigg] \ . 
\end{split}
\ee

\subsection{Orbital and spin evolution}

For the semimajor axis, we have (Eqs.\,(\ref{211110c}) and (\ref{211103c}))
\be
\frac{\dot a}{a} = - \frac{\At}{\beta n  a^2}  \sumk  \frac{k}{2} \left[ \bz  \Xtz   
   - 3 \, \bd    \Xtdp  \right]
\ , \llabel{211103a}
\ee
or, up to $e^2$,
\be
\llabel{221106t4}
\begin{split}
\frac{\dot a}{a} = \frac{3 \At}{\beta n a^2} & \Bigg\{
  \bDwmDn   \left(1-6 e^2\right) 
 - \bZwpUn \frac{1}{2} e^2 \\ &
 + \bDwmUn \frac{1}{4}  e^2 
 + \bDwmTn \frac{63}{4}  e^2
 \Bigg\} \ .
\end{split}
\ee
Since $x=1$, for the eccentricity we get (Eq.\,(\ref{211110e})),
\be
\llabel{211029y}
\begin{split}
\dot e  = - \dfrac{\At}{4} \, \frac{\sq}{\beta n a^2 e} & \sumk  \Bigg\{ \bz 
   \Xtz  k \sq  \\
  & + 3 \, \bd \Xtdp  \left(2-k \sq\right) \Bigg\} \ ,
\end{split}
\ee
or, up to $e^2$,
\be
\llabel{221106te}
\begin{split}
\dot e  =  - \dfrac{3 \At e}{4 \beta n a^2} &  \Bigg\{ 
  \bZwpUn  
 +  \bDwmDn 
 + \frac{1}{2}  \bDwmUn 
 - \frac{21}{2}   \bDwmTn 
 \Bigg\} \ ,
\end{split}
\ee
and for the rotation rate (Eq.\,(\ref{211110t5}))
\be
\dot \om = - \frac{3 \At}{2 C}  \sumk \bd  \Xtdp  
\ , \llabel{211103b}
\ee
or, up to $e^2$,
\be
\llabel{221106tw}
\begin{split}
\dot \om = - \frac{3 \At}{2 C} & \Bigg\{  
 \bDwmDn \left(1-6 e^2\right) \\ &
 + \bDwmUn \frac{1}{2}  e^2 
 + \bDwmTn \frac{21}{2}  e^2
 \Bigg\} \ .
\end{split}
\ee
The evolution of the obliquity is simply given by $\dot \thetao =0$ (Eq.\,(\ref{211110i})) since $\sin \thetao = 0$, that is, the motion remains planar.

\subsection{Energy balance}

The total energy transferred to the planet due to tides is obtained from expression (\ref{211110l}) with $x=1$,
\be
\llabel{211103e}
\begin{split}
\Pt = - \dfrac{\At}{4}  \sumk & \Bigg\{ k n \,\bz \Xtz   \\& 
+ 3 \, (2 \om -k n) \, \bd  \Xtdp  \Bigg\} \ .
\end{split}
\ee

\section{Conclusion}

\llabel{sectconc}

In this paper, we have revisited the spin and orbital evolution of a planet with a dense atmosphere under the action of thermal tides.
We derive the secular equations of motion using a vectorial formalism, where the basis only depends on the unit vectors of the spin and orbital angular momenta.
These vectors are related to the spin and orbital quantities, thus they are easy to obtain and independent of the chosen frame.
The equations only depend on series of Hansen coefficients, which are widely used in celestial mechanics.
They are obtained after averaging over the mean anomaly and over the argument of the pericenter, because thermal tides are expected to modify the orbital elements on a timescale much longer than the evolution of these two angles.
In some problems, where the pericenter evolves slowly, we can also perform a single average over the mean anomaly following the method presented in Sect.~3 of \citet{Correia_Valente_2022} for bodily tides.

The expression of the second harmonic of the surface pressure variations (Eq.\,(\ref{220929a})) was obtained following the approximations in the work by \citet{Auclair-Desrotour_etal_2017a}, which is in very good agreement with the results obtained with general circulation models \citep{Leconte_etal_2015, Auclair-Desrotour_etal_2019a}.
Nevertheless, for more refined atmospheric models \citep[e.g.,][]{Wu_etal_2023}, we also expect that we only need to correct the expression for the surface pressure variations (Eq.\,(\ref{220929a})) and that the dynamical equations derived in Sect.~\ref{secdyn} do not change.

The vectorial formalism presented in this paper is well suited to study the long-term evolution of Earth-mass rocky planets. 
In addition to thermal atmospheric tides, we generally need to consider the bodily tides \citep[e.g.,][]{Correia_Valente_2022}, rotational deformation and general relativity corrections \citep[e.g.,][]{Correia_etal_2011}. 
For multiplanet systems, the secular interactions can be obtained either by developing the perturbing functions in terms of Laplace coefficients, suited for nonresonant compact systems \citep[e.g.,][]{Boue_Fabrycky_2014a}, or in terms of Legendre polynomials, suited for hierarchical systems \citep[e.g.,][]{Correia_etal_2016}.
All these previous studies adopted the same kind of reference vectors and thus each individual effect can be simply added to get the complete dynamical evolution of the planet.

\begin{acknowledgements}
This work was supported by grant
SFRH/BD/137958/2018,
and by projects
CFisUC (UIDB/04564/2020 and UIDP/04564/2020),
GRAVITY (PTDC/FIS-AST/7002/2020),
PHOBOS (POCI-01-0145-FEDER-029932), and
ENGAGE SKA (POCI-01-0145-FEDER-022217),
%Enabling Green E-science for the SKA Research Infrastructure (ENGAGE SKA), 
funded by COMPETE 2020 and FCT, Portugal.
\end{acknowledgements}

\bibliographystyle{aa}
\bibliography{\bibpath correia}

\end{document}